\documentclass{elsart}
\usepackage{epsfig}

\begin{document}


\begin{frontmatter}

\title{Differential cross sections, charge production asymmetry, and
spin-density matrix elements for \boldmath $D^{*\pm}$(2010) produced in
500 GeV/\boldmath $c$ $\pi^-$ - nucleon interactions.}

\collab{Fermilab E791 Collaboration}

\author[inst_10]{E.~M.~Aitala,}
\author[inst_1]{S.~Amato,}
\author[inst_1]{J.~C.~Anjos,}
\author[inst_6]{J.~A.~Appel,}
\author[inst_16]{D.~Ashery,}
\author[inst_6]{S.~Banerjee,}
\author[inst_1]{I.~Bediaga,}
\author[inst_9]{G.~Blaylock,}
\author[inst_17]{S.~B.~Bracker,}
\author[inst_15]{P.~R.~Burchat,}
\author[inst_7]{R.~A.~Burnstein,}
\author[inst_6]{T.~Carter,}
\author[inst_1]{H.~S.~Carvalho,}
\author[inst_14]{N.~K.~Copty,}
\author[inst_10]{L.~M.~Cremaldi,}
\author[inst_20]{C.~Darling,}
\author[inst_6]{K.~Denisenko,}
\author[inst_3]{S.~Devmal,}
\author[inst_12]{A.~Fernandez,}
\author[inst_14]{G.~F.~Fox,}
\author[inst_2]{P.~Gagnon,}
\author[inst_1,inst_13]{C.~Gobel,}
\author[inst_10]{K.~Gounder,}
\author[inst_6]{A.~M.~Halling,}
\author[inst_4]{G.~Herrera,}
\author[inst_16]{G.~Hurvits,}
\author[inst_6]{C.~James,}
\author[inst_7]{P.~A.~Kasper,}
\author[inst_6]{S.~Kwan,}
\author[inst_11]{D.~C.~Langs,}
\author[inst_2]{J.~Leslie,}
\author[inst_6]{B.~Lundberg,}
\author[inst_1]{J.~Magnin,}
\author[inst_1]{A.~Massafferri,}
\author[inst_16]{S.~MayTal-Beck,}
\author[inst_3]{B.~Meadows,}
\author[inst_1]{J.~R.~T.~de~Mello~Neto,}
\author[inst_8]{D.~Mihalcea,}
\author[inst_18]{R.~H.~Milburn,}
\author[inst_1]{J.~M.~de~Miranda,}
\author[inst_18]{A.~Napier,}
\author[inst_8]{A.~Nguyen,}
\author[inst_3,inst_12]{A.~B.~d'Oliveira,}
\author[inst_2]{K.~O'Shaughnessy,}
\author[inst_7]{K.~C.~Peng,}
\author[inst_3]{L.~P.~Perera,}
\author[inst_14]{M.~V.~Purohit,}
\author[inst_10]{B.~Quinn,}
\author[inst_19]{S.~Radeztsky,}
\author[inst_10]{A.~Rafatian,}
\author[inst_8]{N.~W.~Reay,}
\author[inst_10]{J.~J.~Reidy,}
\author[inst_1]{A.~C.~dos Reis,}
\author[inst_7]{H.~A.~Rubin,}
\author[inst_10]{D.~A.~Sanders,}
\author[inst_3]{A.~K.~S.~Santha,}
\author[inst_1]{A.~F.~S.~Santoro,}
\author[inst_3]{A.~J.~Schwartz,}
\author[inst_4,inst_19]{M.~Sheaff,}
\author[inst_8]{R.~A.~Sidwell,}
\author[inst_20]{A.~J.~Slaughter,}
\author[inst_3]{M.~D.~Sokoloff,}
\author[inst_1,inst_5]{C.~J.~Solano Salinas,}
\author[inst_8]{N.~R.~Stanton,}
\author[inst_6]{R.~J.~Stefanski,}
\author[inst_19]{K.~Stenson}                                                   
\author[inst_10]{D.~J.~Summers,}
\author[inst_20]{S.~Takach,}
\author[inst_6]{K.~Thorne,}
\author[inst_8]{A.~K.~Tripathi,}
\author[inst_19]{S.~Watanabe,}
\author[inst_16]{R.~Weiss-Babai,}
\author[inst_11]{J.~Wiener,}
\author[inst_8]{N.~Witchey,}
\author[inst_20]{E.~Wolin,}
\author[inst_8]{S.~M.~Yang,}
\author[inst_10]{D.~Yi,}
\author[inst_8]{S.~Yoshida,}
\author[inst_15]{R.~Zaliznyak,}
\author[inst_8]{and C.~Zhang}

\address[inst_1]{Centro Brasileiro de Pesquisas F\'\i sicas, 
                 Rio de Janeiro RJ, Brazil}
\address[inst_2]{University of California, Santa Cruz, California 95064}
\address[inst_3]{University of Cincinnati, Cincinnati, Ohio 45221}
\address[inst_4]{CINVESTAV, 07000 Mexico City, DF Mexico}
\address[inst_5]{Universidade Federal de Itajub\'a,
                 Itajub\'a, Brazil}
\address[inst_6]{Fermilab, Batavia, Illinois 60510}
\address[inst_7]{Illinois Institute of Technology, Chicago, Illinois 60616}
\address[inst_8]{Kansas State University, Manhattan, Kansas 66506}
\address[inst_9]{University of Massachusetts, Amherst, Massachusetts 01003}
\address[inst_10]{University of Mississippi, University, Mississippi 38677}
\address[inst_11]{Princeton University, Princeton, New Jersey 08544}
\address[inst_12]{Universidad Autonoma de Puebla, Puebla, Mexico}
\address[inst_13]{Universidad de la Rep\'ublica, Montevideo, Uruguay}
\address[inst_14]{University of South Carolina, Columbia, South Carolina 29208}
\address[inst_15]{Stanford University, Stanford, California 94305}
\address[inst_16]{Tel Aviv University, Tel Aviv, 69978 Israel}
\address[inst_17]{Box 1290, Enderby, British Columbia, V0E 1V0, Canada}
\address[inst_18]{Tufts University, Medford, Massachusetts 02155}
\address[inst_19]{University of Wisconsin, Madison, Wisconsin 53706}
\address[inst_20]{Yale University, New Haven, Connecticut 06511}


\begin{abstract}

We report differential cross sections for the production of
$D^{*\pm}$(2010) produced in 500 GeV/$c$ $\pi^-$-nucleon interactions
from experiment E791 at Fermilab, as functions of Feynman-$x$ ($x_F$)
and transverse momentum squared ($p_T^2$).  We also report the
$D^{*\pm}$ charge asymmetry and spin-density matrix elements as
functions of these variables.  Investigation of the spin-density
matrix elements shows no evidence of polarization. The average values
of the spin alignment are $\langle\eta\rangle = 0.01\pm 0.02$ and
$-0.01\pm 0.02$ for leading and non-leading particles, respectively.

\end{abstract}

\begin{keyword}
PACS 13.85.Ni, 13.88.+e, 14.40.Lb;
charm hadroproduction;
differential cross section;
spin alignment;
polarization
\end{keyword}

\end{frontmatter}

\section{Introduction}

$D^{*\pm}$(2010) ($J^P=1^-$) production is important to understanding
the production of charm because $D^*$'s are expected to dominate the
charm cross-section: their spin causes them to be favored threefold
over the low-lying $D$ mesons at high center of mass energies. Charm
production is an interesting topic in its own right, being calculable
in perturbative QCD\cite{ref-nloqcd}. There has also been interest in
the intrinsic charm content of hadrons\cite{ref-intr}. Finally,
accurate simulations of charm production required by future
experiments at the Tevatron, the LHC, and lepton colliders will
depend on these measurements which can be used to tune {\sc
Pythia}\cite{ref-pythia} and other Monte Carlo simulation packages.

Since $D^*$ mesons have spin~1, we might expect the charm quark spin
to be reflected in the meson. Spin retention is particularly strong
for heavy quarks, where the usual stiffness of the spin vector under
relativistic boosts may be expected to play a part\cite{ref-bmt}. In
any case, there are many anomalies of spin and polarization. For
instance, it is known that some hyperons are strongly polarized in
some regions of production and not in others, the reasons for this
behavior being still not completely understood\cite{ref-hyppol}. It is
also known that most of the spin of the proton is not carried by the
valence quarks\cite{ref-psf}.  Copious production of a vector meson
provides us with a unique opportunity to add new information on
polarization, and we hope that these data will shed further light on
hadron physics.

In this paper, we report measurements of the production of
$D^{*\pm}$(2010) in 500~GeV/$c$ $\pi^-$-nucleon interactions.
Forward differential cross sections and $D^{*\pm}$ production
asymmetries have been measured as functions of Feynman-x ($x_F$) and
transverse momentum squared ($p_T^2$).  The measurements are compared
to {\sc Pythia} Monte Carlo models.  The differential distributions
are fit to well-known functional forms and fit parameters are compared
with previous measurements.  Also, spin-density matrix elements were
measured, again as functions of $x_F$ and $p_T^2$.


\section{Experiment and Data Sample}
E791, a high statistics charm physics experiment, took data at
Fermilab's Tagged Photon Laboratory during
the 1991--2 fixed-target run.  The
experiment~\cite{ref-detect} used an upgraded version of the
two-magnet spectrometer previously used in E516, E691, and E769.
A 500~GeV/$c$ $\pi^-$ beam was directed at five target foils: a
0.52~mm thick platinum foil, followed by four 1.56~mm thick carbon
foils, with a typical spacing of 15~mm between foil centers.  Tracks
and vertices used hits in 23 silicon microstrip and 45 wire chamber
planes.  The spectrometer included two \v{C}erenkov counters,
electromagnetic and hadronic calorimeters, and two muon scintillator
walls.  A loose trigger based on transverse
calorimeter~\cite{ref-trigger} energy was employed.  An innovative data
acquisition system~\cite{ref-da} recorded $2\times 10^{10}$ events.  The
resulting 50 Terabyte data set was reconstructed using large computing
farms at four sites~\cite{ref-farms}.

The $D^*$ sample used in this analysis was collected from the decay
mode $D^{*+}\rightarrow D^0\pi^+$.\footnote{Charge conjugates are
implied for all decay modes discussed in this paper.}  The $D^0$
candidates were obtained in the decay modes $D^0\rightarrow K^-\pi^+$
and $D^0\rightarrow K^-\pi^-\pi^+\pi^+$.  Selection criteria were
chosen to optimize $S/\sqrt{S+B}$, where $S$ and $B$ are respectively
the signal and background in $D^{*\pm}$ $Q$-value histograms.  Here, the
$D^{*\pm}$ $Q$-value is defined as $m(D^*) - m(D^0) - m(\pi^+)$ where
$m(D^*)$ is the reconstructed $D^{*\pm}$ mass, $m(D^0)$ is the
reconstructed $D^0$ mass, and $m(\pi^+)$ is the charged pion
mass~\cite{ref-pdb}.  Owing to the high statistics of the sample and
the small number of criteria used for selection, the data set was used
directly for this optimization.

\begin{figure}
\centerline{\epsfig{figure=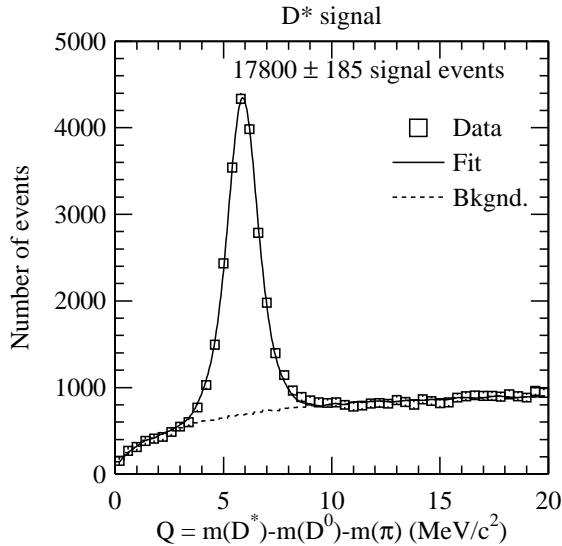,height=3in}}
\caption{$D^{*\pm}$ sample passing selection criteria used in this
analysis.  In the fit shown, a double Gaussian was used to model the
signal shape and random event mixing was used to model the background.
This fit yields $17800\pm 185$ $D^{*\pm}$ signal events.}
\label{fig-all3al}
\end{figure} 

The short lifetime of the $D^{*\pm}$ makes it impossible to
reconstruct a $D^{*\pm}$ decay vertex separate from the primary
vertex, so in what follows, the secondary vertex is defined to be the
point where the $D^0$ decays, and the primary vertex to be the point
where the $D^{*\pm}$ and $D^0$ are produced.  The longitudinal
separation between the primary and secondary vertices was required to
be at least $8\:(7)$ standard deviations\footnote{In some instances,
different cuts were applied for the $K\pi$ and $K\pi\pi\pi$ modes.  In
these cases the $K\pi\pi\pi$ cut is shown in parentheses.}, and the
secondary vertex was required to be outside any target foil.  The
component of the momentum vector of the $D^0$ candidate perpendicular
to the line joining the primary and secondary vertices was required to
be less than $0.4\:(0.45)\ {\rm GeV}/c$, and also the $D^0$ momentum
vector was required to point back to within 0.1~mm of the primary
vertex in the plane transverse to the beam.  To reduce background
caused by false secondary vertices or tracks being mis-assigned to the
secondary vertex, it was further required that the product over all
decay tracks of the ratio of the transverse impact parameter at the
secondary vertex to that at the primary vertex be less than one.  To
further reduce background, the reconstructed mass of the $D^0$
candidate was required to fall within a mass window which was
determined by creating histograms of the masses of the $D^0$
candidates separately in the various regions of $D^{*\pm}$ $x_F$ and
$p_T^2$ to which a fit was performed where the $D^0$ signal was
represented by a Gaussian and the background by a straight line.  The
limits of the mass window were chosen to be at $2.5\sigma$ on either
side of the mean.  Owing to the small phase space available to the
$D^{*\pm}$ decay, the background in the $Q$-value histogram is already
low enough that no additional cuts on the $D^{*\pm}$ decay pion
candidates were necessary, other than basic track quality requirements
applied to all decay tracks.  Applying the above selection criteria
yields a data sample of $17800\pm 185$ $D^{*\pm}$ signal events which
are shown in Figure~\ref{fig-all3al}.

\section{Differential Production Cross Sections}

For the differential cross section study, the data sample was divided
into 12 bins each of $x_F$ and $p_T^2$.  In the range $-0.1 < x_F <
0.4$, the bins have a width of 0.05, and in the range $0.4 < x_F <
0.6$ the bins have a width of 0.1.  In the range $0 < p_T^2 < 2\;({\rm
GeV}/c)^2$ bins have a width of 0.5~$({\rm GeV}/c)^2$, and in the
range $2 < p_T^2 < 10\; ({\rm GeV}/c)^2$ bins have a width of 1~$({\rm
GeV}/c)^2$.  In each bin of $x_F$ or $p_T^2$, $D^{*\pm}$ $Q$-value
histograms were created.  Random event mixing was used to model the
background shape, and the background level was determined by fitting
the event mixed histogram to the corresponding data histogram using a
binned maximum likelihood fit in the background region ($10 < Q <
20\;{\rm MeV}/c^2$). In the signal region ($0 < Q < 10\; {\rm
MeV}/c^2$) we subtract from the data histogram the scaled number of
events from the event-mixed histogram to obtain the number of signal
events.  In this way, it isn't necessary to model the details of the
signal shape to determine the number of signal events in the
histogram.

To model the acceptance, 14 million Monte Carlo events were generated
which, when reconstructed and subjected to the same selection criteria
as our data, yielded approximately ten times as many $D^{*\pm}$ events as
in the data sample.  The Monte Carlo samples were fit in the same way as
the data to obtain a bin-by-bin efficiency estimate.

\begin{figure}
\centerline{\epsfig{figure=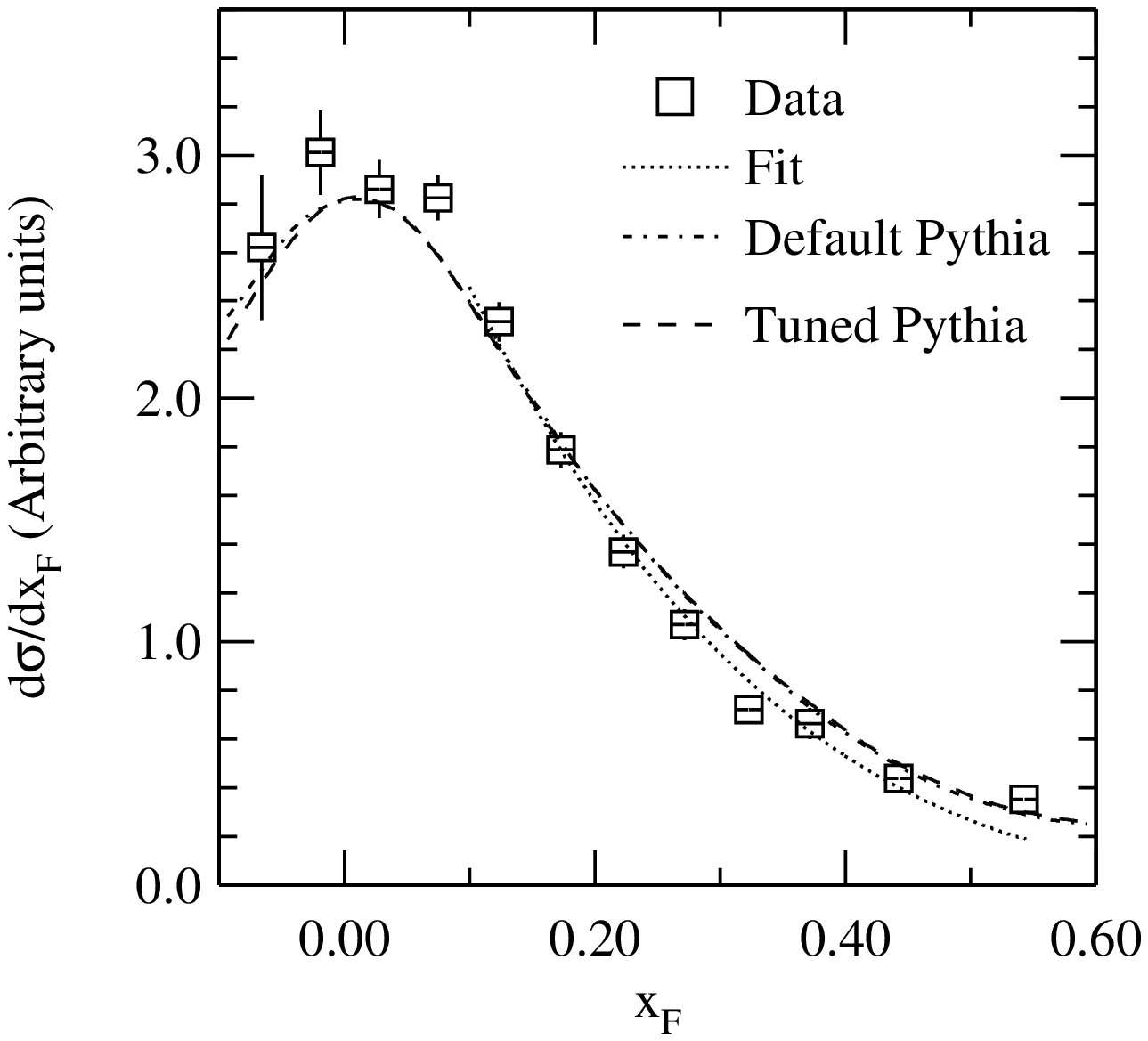,height=3in}\epsfig{figure=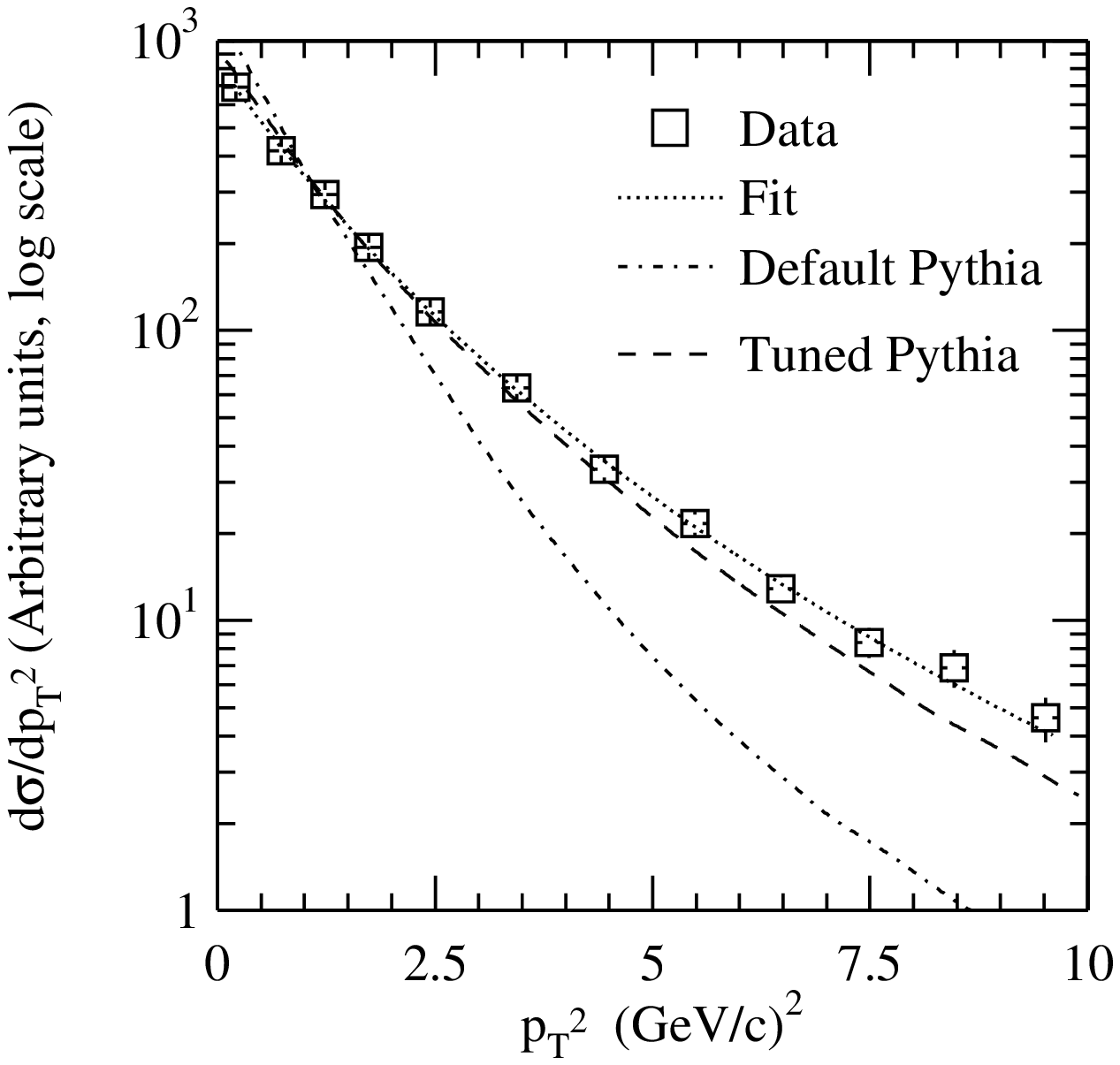,height=3in}}
\caption{Acceptance-corrected differential cross sections of $D^{*\pm}$
production as functions of $x_F$ (left) and $p_T^2$ (right).  Error
bars include both statistical and systematic errors.  Fits shown are
of functional forms described in Equations \protect{\ref{eq-xf}} and
\protect{\ref{eq-ptmangano}}.  Also, default and $D^\pm$-tuned {\sc
Pythia}
models are plotted.}
\label{fig-formfits}
\end{figure} 

The acceptance-corrected yields were used to plot the differential
cross sections in Figure~\ref{fig-formfits}.  For the $x_F$
distributions, we used data in the range $0 < p_T^2<10$~GeV$^2$, and
for the $p_T^2$ distributions, $-0.1 < x_F < 0.6$.  The $x_F$
distribution was fit in the range 0.1--0.6 with the traditional
function:
\begin{equation}
\frac{d\sigma}{dx_F}\sim(1-x_F)^n
\label{eq-xf}
\end{equation}
The $p_T^2$
distribution was fit with the traditional function
\begin{equation}
\frac{d\sigma}{dp_T^2}\sim\exp(-Bp_T^2)
\label{eq-pt}
\end{equation}
in the range $0 < p_T^2 < 4\;({\rm GeV}/c)^2$ and with the function
\begin{equation}
\frac{d\sigma}{dp_T^2}\sim\left(\frac{1}{bmc^2+p_T^2}\right)^\beta
\label{eq-ptmangano}
\end{equation}
in the full range $0 < p_T^2 < 10\;({\rm GeV}/c)^2$.  This last
functional form was suggested by Frixione et al.\cite{ref-mangano}.

In Figure~\ref{fig-formfits}, the E791 results are also compared to
the default {\sc Pythia} model~\cite{ref-pythia} and a ``tuned'' {\sc
Pythia} model~\cite{ref-tuned} in which parameters are adjusted to
more closely model the $D^{\pm}$ production characteristics reported
in an earlier E791 paper~\cite{ref-carter}.  The results of the
various fits are listed in Table~\ref{tab-diff}, along with results
from previous
experiments~\cite{ref-wa92,ref-769,ref-na32b,ref-na32a,ref-na27}.

\begin{table}
\caption{Parameters from fits to $\pi^{\pm}N$ differential $D^{*\pm}$
cross sections vs. $x_F$ and $p_T^2$, including comparisons to other
experiments (\cite{ref-wa92,ref-769,ref-na32b,ref-na32a,ref-na27}).
Column labels refer to fit parameters of the same name defined by
Equations~\protect{\ref{eq-xf}}--\protect{\ref{eq-ptmangano}}.  The
labels ``l'' and ``nl'' refer to leading and non-leading particles,
respectively. Asterisks (*) indicate extended range of $p_T^2$ used in
fit of Equation~\ref{eq-ptmangano}.  Errors shown on E791 results
include both statistical and systematic components.}
\begin{tabular}{cccccccc} 
\hline\hline
Expt. & \parbox{0.4in}{\begin{center}$P_{\rm beam}$ \\ GeV/$c$ \end{center}} & \parbox{0.5in}{\begin{center}$x_F$ fit \\ range \end{center}} & $n$ & \parbox{0.67in}{\begin{center}$p_T^2$~fit~range \\ (Gev/c)$^2$\end{center}} & $B$ & \parbox{0.6in}{\begin{center}{$bmc^2$ *}\\ $({\rm GeV}/c)^2$\end{center}} & $\beta$ * \\ \hline
E791 & 500 & 0.1 to 0.6 & $3.8 \pm 0.2$ & 0 to 4 & $0.75 \pm 0.02$ & $5.1 \pm 0.8$ & $5.0 \pm 0.6$ \\ 
l.~($D^{*-}$) & & & $3.6 \pm 0.2$ & or & $0.73 \pm 0.02$ & $6.1 \pm 1.5$ & $5.6 \pm 1.0 $ \\ 
nl.~($D^{*+}$) & & & $4.1 \pm 0.3$ & 0 to 10* & $0.77 \pm 0.03$ & $4.0 \pm 1.1$ & $4.4 \pm 0.8$ \\ \hline
WA92\cite{ref-wa92} & 350 & 0 to 0.6 & $4.3\pm0.4$ & 0 to 4 & $0.84\pm0.05$ & $3.6\pm 0.8$ & $4.7\pm 0.7$ \\
l. & & & $4.9\pm 0.5$ & or & & & \\
nl. & & & $3.9\pm 0.4$ & 0 to 14* & & & \\ \hline
E769\cite{ref-769} & 250 & 0.1 to 0.6 & $3.5\pm 0.3$ & 0 to 4 & $0.70\pm 0.07$ &  &  \\ 
l. & & & $2.9\pm 0.4$ & & $0.58\pm 0.09$ &  &  \\ 
nl. & & & $4.1\pm 0.5$ & & $0.79\pm 0.09$ &  &  \\ \hline
NA32\cite{ref-na32b} & 230 & 0 to 0.8 & $3.14^{+0.40}_{-0.39}$ & 0 to 10 & $0.79\pm 0.07$ & & \\ 
l. & & & $2.62^{+0.53}_{-0.49}$ & & $0.71^{+0.09}_{-0.08}$ & & \\
nl. & & & $3.83^{+0.66}_{-0.62}$ & & $0.90\pm 0.11$ & & \\ \hline
NA32\cite{ref-na32a} & 200 & 0 to 0.7 & $2.8^{+1.1}_{-0.9}$ & 0 to 5 & $0.9^{+0.3}_{-0.2}$ & & \\
l. & & & $4.7^{+1.9}_{-1.6}$ & & & & \\
nl. & & & $1.7^{+1.4}_{-1.0}$ & & & & \\ \hline
NA27\cite{ref-na27} & 360 & 0 to 0.5 & $4.3^{+1.8}_{-1.5}$ & 0 to 3 & $0.9\pm 0.4$ & & \\
\hline\hline
\end{tabular}
\label{tab-diff}
\end{table}

\section{Charge Asymmetry in Production}

The charge production asymmetry is defined by the parameter:
\begin{equation}
A\equiv \frac{\sigma(D^{*-})-\sigma(D^{*+})}{\sigma(D^{*-})+\sigma(D^{*+})} 
 \approx\frac{N(D^{*-})-N(D^{*+})}{N(D^{*-})+N(D^{*+})}
\label{eq-asym}
\end{equation}
where $\sigma(D^{*-})$ and $\sigma(D^{*+})$ denote the production
cross-sections, of $D^{*-}$ and $D^{*+}$, respectively, and 
$N(D^{*-})$ and $N(D^{*+})$ denote the respective acceptance-corrected
numbers of such particles observed.

\begin{figure}
\centerline{\epsfig{figure=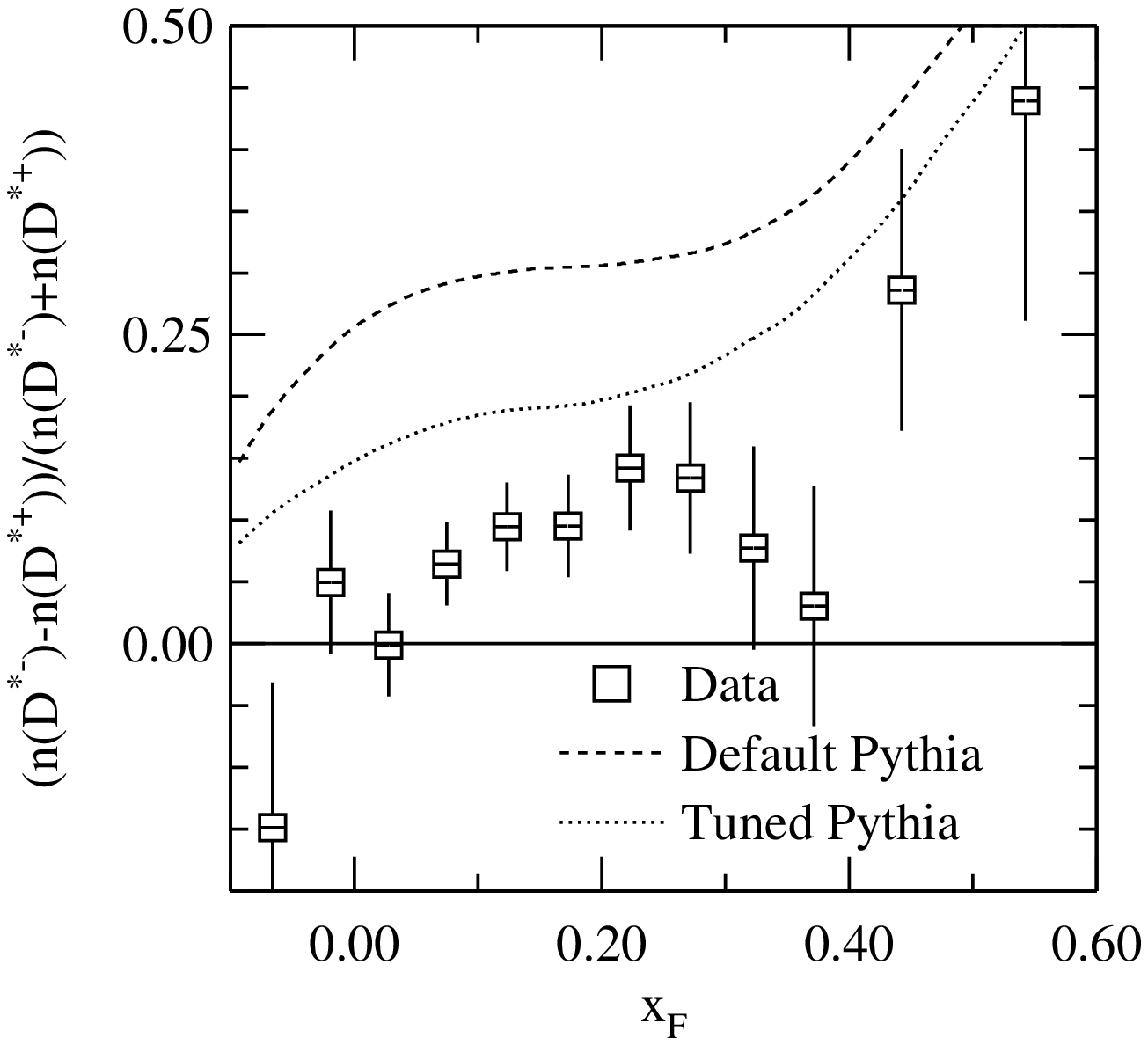,height=3in}\epsfig{figure=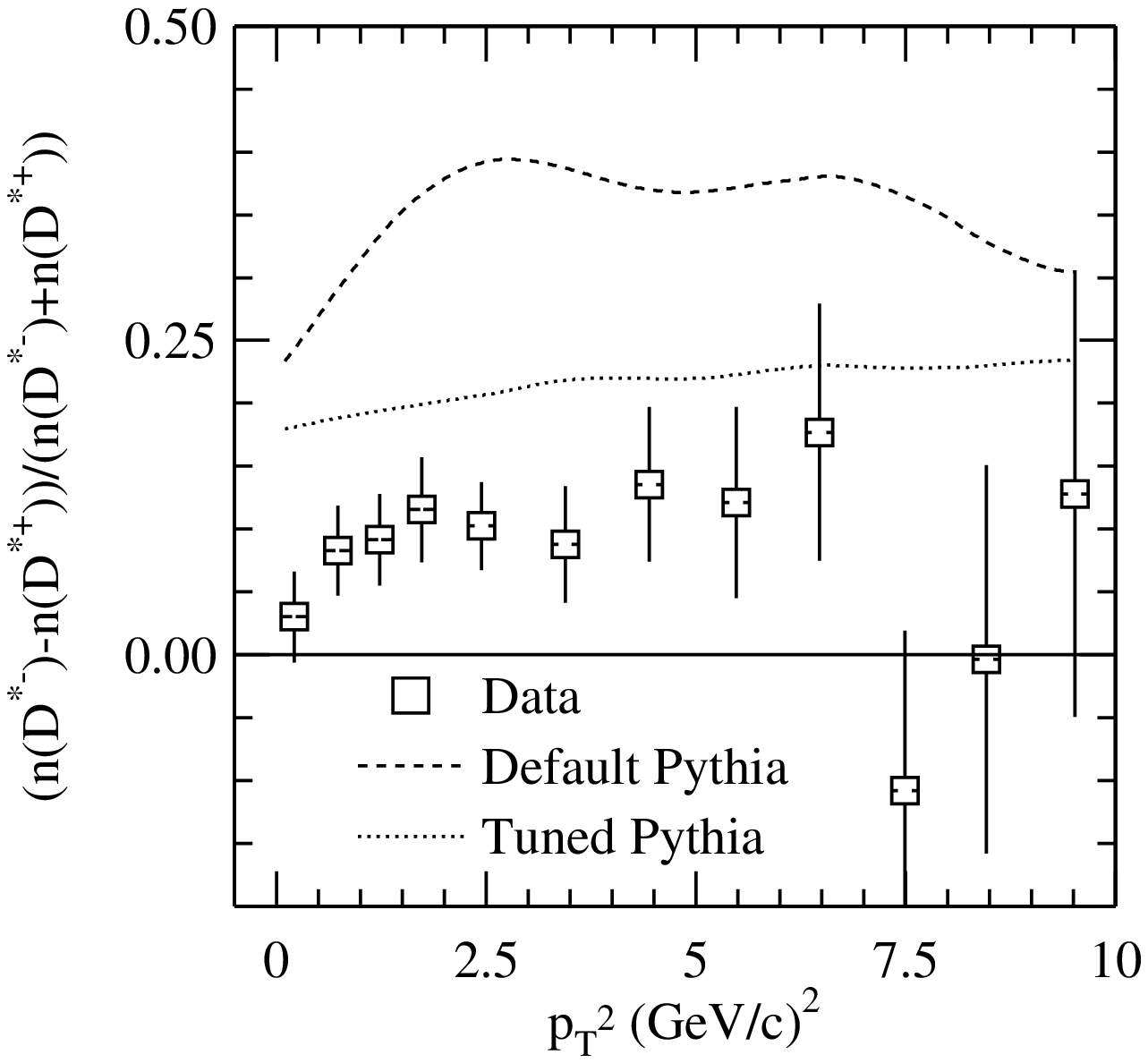,height=3in}}
\caption{Charge production asymmetry as functions of $x_F$ and
$p_T^2$.  (Error bars include statistical and systematic errors added
in quadrature.  Horizontal error bars indicate RMS values of $x_F$ or
$p_T^2$ distributions within the bins.)  Comparisons to default and
$D^\pm$-tuned {\sc Pythia} $D^{*\pm}$ asymmetry predictions also shown.}
\label{fig-asym}
\end{figure}

Using the same fitting techniques and data sample as was used in the
differential cross section study, the production asymmetry as
functions of $x_F$ and $p_T^2$ are shown in Figure~\ref{fig-asym}.

\section{Spin-Density Matrix Elements}

The polarization state of a spin 1 particle can be described by a
complex $3\times 3$ spin-density matrix.  A complex $3\times 3$ matrix
has 18 real components, but hermiticity and the fact that ${\rm
Tr}\rho=1$ reduces this number to 8.  Also, the $D^{*\pm}$'s are
produced in a strong interaction where parity conservation further
reduces this number to 4.  The spin-density matrix when expressed in a
helicity basis then takes this form~\cite{ref-perl}:

\[\rho = \left( \begin{array}{ccc}
           \rho_{11} & \rho_{10} & \rho_{1-1} \\
           \rho_{10}^* & 1-2\rho_{11} & -\rho_{10}^* \\
           \rho_{1-1} & -\rho_{10} & \rho_{11} \end{array} \right) \]

with $\rho_{1-1}$ and of course $\rho_{11}$ real.

From this, one can derive the angular distribution~\cite{ref-perl}, 

\begin{eqnarray}
\frac{d\sigma}{d\Omega} \sim  \frac{1}{4\pi}[ & 1+(1-3\cos^2\theta)(3\rho_{11}-1) \nonumber \\
 &  -(3\sqrt{2}\sin2\theta\cos\phi)\Re\rho_{10} \nonumber \\
 &  -(3\sin^2\theta\cos2\phi)
\rho_{1-1} ].\label{eq-dist}
\end{eqnarray}

The angles $\theta$ and $\phi$ in the above equation are polar and
azimuthal angles respectively in a right-handed spherical polar
coordinate system in the rest frame of the $D^{*\pm}$, whose polar axis
($\theta=0$) points along the $D^*$ boost vector (a ``helicity''
basis, with the $z$-axis chosen to lie in the production plane) and
the $\phi=0, \theta=\pi/2$ axis is chosen to point in the direction
${\bf \hat{d}} \times {\bf \hat{b}}$ where ${\bf \hat{d}}$ is a unit
vector in the direction of the $D^*$ momentum vector and ${\bf
\hat{b}}$ is a vector in the direction of the beam particle momentum
vector (the $x$-axis is normal to the production plane).

Note that this distribution depends on only three of the four
independent components of the spin-density matrix:
$\rho_{11}$, $\Re\rho_{10}$, and $\rho_{1-1}$.  Thus $\Im\rho_{10}$
cannot be determined from this angular distribution study.

The degree of polarization is commonly expressed by the spin alignment
parameter:
\[ \eta = 1-3\rho_{11} \]

To measure the spin-density matrix elements, the data sample was first
divided into five bins each of $x_F$ and $p_T^2$.  In each of these
bins, an $8\times 8$ two-dimensional histogram of $d\sigma/d\Omega$
vs. $(\cos\theta, \phi)$ was created.  The $D^{*\pm}$ yield in each bin
was measured as follows.  Events were counted both above and below a
$Q$-value of 10~${\rm MeV}/c^2$ for both the data histogram and the
event-mixed data histogram (our background model).  A comparison of
the number of events in the data and mixed data sample in the
$>10\;{\rm MeV}/c^2$ region yields a scale factor, which is used in
subtracting the number of background events from the signal region
$<10\;{\rm MeV}/c^2$.  The same procedure was applied to a Monte Carlo
sample to obtain a binned acceptance function, and the data sample
was corrected for acceptance in this way.

\begin{figure}
\centerline{\epsfig{figure=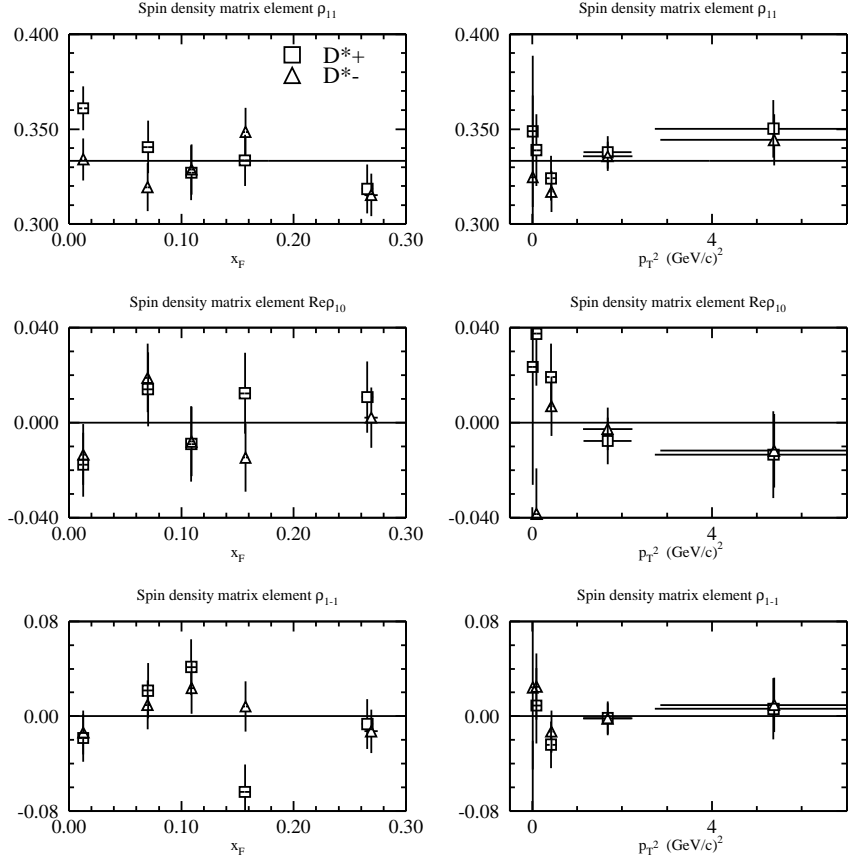,height=4.5in}}
\caption{Spin-density matrix elements as functions of $x_F$ and
$p_T^2$.  (Error bars include systematic as well as statistical errors
added in quadrature.)  The solid line corresponds to unpolarized
$D^*$'s.}
\label{fig-spdm}
\end{figure}

The resulting two dimensional histogram was fitted with the functional
form shown in Equation~\ref{eq-dist} to obtain values for the
spin-density matrix elements.  These results are plotted separately
for $D^{*+}$ and $D^{*-}$ in Figure~\ref{fig-spdm}.  Because the
sample had to be split into sub-regions of $\cos\theta$ and $\phi$ in
order to perform the fits, fewer bins in $x_F$ and $p_T^2$ were used
than in previous sections.  There is no measurable polarization at our
level of sensitivity in any sub-region of the $x_F$ or $p_T^2$ range
examined.  The average value of the spin alignment parameter is found
to be $\langle\eta\rangle=0.01\pm0.02$ and $-0.01\pm0.02$ for leading
($D^{*-}$) and non-leading ($D^{*+}$) particles respectively.

\section{Systematic Errors}

The systematic errors reported were determined by comparing results,
either from various subsamples, or by analyzing the full sample in
multiple ways.  Differences in measurements of the number of observed
$D^{*\pm}$ signal events were then determined as fractions of
statistical errors bin by bin.  Root mean squares of these fractions
were then used as estimates of the systematic error as a fraction of
statistical error.  Several possible sources of systematic error were
considered, and those found to be significant are tabulated in
Table~\ref{tab-systerr}.  The cut on the $D^0$ mass varies with $x_F$
and $p_T^2$ and depends on our modeling of the $D^0$ signal.  By
comparing results with different $D^0$ mass cuts, a systematic error
of 24\% (of the statistical error) was assigned for this effect.
Because the default {\sc Pythia} $p_T^2$ and $x_F$ distributions don't
match those in data exactly, it was necessary to weight the Monte
Carlo events to correct the distributions when modeling acceptance as
a function of the other variable.  By varying the ``weighting curve''
parameters by $1\sigma$ from their central values, it was determined
that a relative systematic error of 13\% should be assigned.  Two
decay modes of the $D^0$ ($K\pi$ and $K\pi\pi\pi$) are used in this
analysis.  Since one mode involves two more decay particles than the
other, one can compare results from the two modes to obtain an
estimate of the error resulting from modeling of the tracking
efficiency.  A relative error of 25\% was assigned for this effect,
although some uncertainty remains as the two samples were
statistically independent, and it is not possible to unambiguously
untangle the systematic and statistical errors.  Event mixing was used
to model the backgrounds in the $D^{*\pm}$ histograms.  If the
event-mixed histograms do not perfectly model the actual background,
this would result in a systematic error.  By varying the range of
$Q$-values in which the background is fit, an estimate of the
magnitude of the systematic error is obtained.  A relative error of
39\% is assigned to this effect.  Over the period of E791's run, the
500~${\rm GeV}/c\;\pi^-$ beam caused a significant amount of
ionization in regions of the drift chambers through which it passed.
This caused deposits to gradually form on the drift chamber wires
resulting in a loss of tracking efficiency in regions through which
the beam passed.  These drift chamber ``holes'' changed in size and
shape throughout the run period.  Although this is modeled in the
Monte Carlo in a time dependent fashion, this modeling is not perfect.
By comparing results obtained from subsamples of various run periods,
a 58\% relative systematic error was assigned to this effect.

Each source of error was measured as a fraction of the statistical
error, and a quadratic sum of the various sources is shown at the end
of the table to be 79\% of the statistical error.  Combining the
statistical with the overall systematic error by adding them in
quadrature results in a combined error which is 1.3 times the
statistical error alone.  These combined errors are used to create the
error bars in the plots in this paper.

\begin{table}
\caption{Summary of systematic errors reported as fractions of
statistical errors.}
\begin{tabular}{lc} 
\hline\hline
Source of Uncertainty & Syst.~/ stat.~err. \\\hline
$p_T^2$/$x_F$ weighting & 13\% \\ 
$D^0$ mass cut & 24\% \\ 
Tracking efficiency & 25\% \\ 
$D^*$ fitting & 39\% \\ 
Drift chamber efficiency modeling & 58\% \\ \hline
Total & 79\% \\ 
\hline\hline
\end{tabular}
\label{tab-systerr}
\end{table}

\section{Conclusions}

We have measured the differential cross sections of $D^{*\pm}$
production as functions of $x_F$ and $p_T^2$.  While the usual
functional forms used by other experiments to parameterize their data
(Equations \ref{eq-xf} and \ref{eq-pt}) do not describe our data well,
Equation~\ref{eq-ptmangano} suggested by Frixione et al. provides an
excellent fit over our full range of $p_T^2$ ($\chi^2 = 8.4$ for 9
degrees of freedom).  The {\sc Pythia} model tuned for the E791
analysis of $D^{\pm}$ production does describe the $p_T^2$ cross
section much better than the default {\sc Pythia} model.

We observe an overall charge asymmetry which favors the leading
particle as predicted by the {\sc Pythia} model, although our overall
observed asymmetry is significantly less than predicted.  We do
observe an increase in asymmetry with $x_F$ as predicted by {\sc
Pythia} and consistent with that for $D^{\pm}$
production\cite{ref-carter}.  The asymmetry is approximately constant
as a function of $p_T^2$; while it dips to zero near $p_T^2=0$, data
points in this region are also consistent with the average value.

We observe no $D^{*\pm}$ polarization at our level of sensitivity,
either on average or as functions of $x_F$ or $p_T^2$.

\begin{ack}
We gratefully acknowledge the assistance of the staffs of Fermilab and
of all the participating institutions. This research was supported by
the Brazilian Conselho Nacional de Desenvolvimento Cient\'\i fico e
Technol\'{o}gio, CONACyT (Mexico), FAPEMIG (Brazil), the Israeli
Academy of Sciences and Humanities, the U.S. Department of Energy, the
U.S.--Israel Binational Science Foundation and the U.S. National
Science Foundation. Fermilab is operated by the Universities Research
Association, Inc., under contract with the United States Department of
Energy.
\end{ack}


\begin{thebibliography}{99}
\bibitem{ref-tuned}
The ``tuned'' parameters were obtained from~\cite{ref-carter}:
PMAS(4,1) (the charm quark mass in GeV/$c^2$) was changed from 1.35 to
1.7, PARP(91) (the average $k_t^2$ in $({\rm GeV}/c)^2$) increased
from 0.44 to 1.0, PARP(93) (the maximum allowable $k_t$) in GeV/$c$)
increased from 2.0 to 5.0, and MSTP(92) the remnant quark-diquark
energy splitting function) changed from 4 to 3.
\end{thebibliography}
\end{document}